\documentstyle[editedvolume,numreferences,epsfig]{crckapb} 

\newcommand{\Kbar}{\not{\!K}}

\newcommand{\Hslash}{\not{\!H}}

\newcommand{\Pbar}{\not{\!P}}

\newcommand{\be}{\begin{equation}}
\newcommand{\ee}{\end{equation}}
\newcommand{\ba}{\begin{eqnarray}}
\newcommand{\ea}{\end{eqnarray}}

\newcommand{\nh}{{\bf      h}}

\newcommand{\nk}{{\bf      k}}
\newcommand{\np}{{\bf      p}}       
\newcommand{\nq}{{\bf      q}}

\begin{opening}
\title{RELATIVISTIC EFFECTS IN QUASIELASTIC ELECTRON SCATTERING}
\author{MARIA B. BARBARO}
\institute{Dipartimento di Fisica Teorica - Universit\`a di Torino and INFN,
           Via P. Giuria 1, 10125 Turin, Italy}
\end{opening}
\begin{document}

Electron scattering is known to be one of the most powerful means to
study both the structure of nuclei and the internal structure of the
nucleon, especially the less known strange and axial form factors.
In particular, inclusive (e,e$'$) processes at or near quasielastic 
peak kinematics have  attracted attention in the last two decades and 
several experiments have been performed with the aim of disentangling the 
longitudinal and transverse contributions to the quasielastic cross section.
These are linked to the hadronic tensor
\begin{equation}
W^{\mu\nu}=\overline{\sum_i}\sum_f\langle f|\hat{J}^\mu |i\rangle^\ast
\langle f|\hat{J}^\nu |i\rangle \delta(E_i+\omega-E_f)
\label{eq1}
\end{equation}
via the relations
\begin{eqnarray}
R^L(q,\omega)&=&\left(\frac{q^2}{Q^2}\right)^2\left[
W^{00}-\frac{\omega}{q}(W^{03}+W^{30})+\frac{\omega^2}{q^2}W^{33}
\right] \\
R^T(q,\omega)&=&W^{11}+W^{22} \ ,
\end{eqnarray}
where $Q_\mu=(\omega,\nq)$ is the four-momentum carried by the virtual photon,
$\hat{J}^\mu$ is the nuclear many-body current operator
and the nuclear states $|i\rangle$ and $|f\rangle$ are exact
eigenstates of the nuclear Hamiltonian with definite four-momentum.
The general form (\ref{eq1}) includes all possible final states that
can be reached through the action of the current operator
$\hat{J}^\mu$ on the exact ground state; here we focus on the 
one-particle one-hole (1p-1h) excitations.

The simple impulse approximation, while explaining the 
electron scattering reaction mechanism around the quasielastic 
peak reasonably well, 
is unable to account for the observed strength in the dip 
region between the quasielastic and $\Delta$ peaks, where meson 
production, including via the $\Delta$, and two-body
currents should be taken into account. 

Most of the existing calculations that include nuclear correlations
have been performed within a non-relativistic
framework, where non-relativistic wave functions
and  current operators are  
obtained using standard expansions which view both the 
momentum transfer $q$ and the energy transfer $\omega$ as
being small compared to the nucleon mass $m$.
For high-energy conditions the current operators 
so obtained are obviously inadequate and new expressions are needed.
This provided the focus for the developments in~\cite{Ama98,Ama99,Alv01}, 
where new current operators that are exact as far as the variables
$\omega$ and $q$ are concerned were derived.
\begin{figure}[h]
\begin{center}
\leavevmode
\epsfig{file=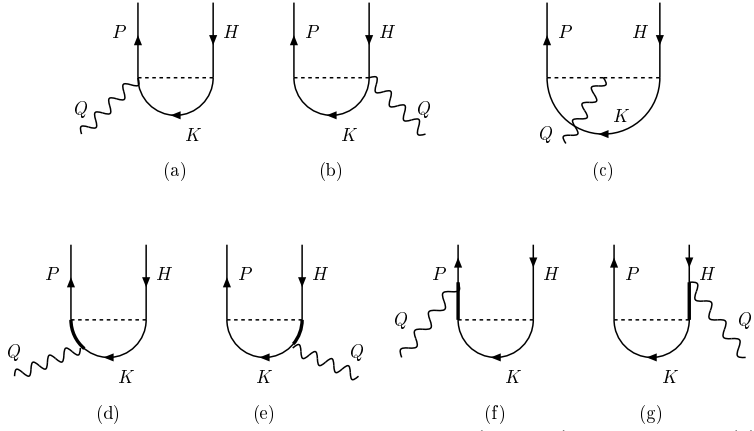,width=10cm}
\end{center} 
\caption{Feynman diagrams representing the seagull (a and b), pion-in-flight
(c), vertex (d and e) and self-energy of the hole (f) and particle (g)
particle-hole matrix elements.}
\end{figure}

Since the physics of the QEP is remote not only 
from the physical surface of the nucleus, but from the Fermi surface as well,
we consider a treatment in terms of nucleonic and mesonic degrees of freedom
(the latter viewed both as force and current carriers) to be appropriate.
Hence, as a first step, we focus on pions, as they can be expected under
these conditions to be the major carrier of the currents that respond to an 
external electromagnetic field impinging on the nucleus.
Moreover, since the effects of pions on the free responses 
of the relativistic Fermi gas are not expected to be too disruptive,
the interaction is dealt with in first order of perturbation
theory, which amounts to consider the many-body Feynman diagrams 
in Fig.~1: they represent all the possible processes involving one and only one
pionic line and connecting the Fermi sphere $|F\rangle$ 
to a 1p-1h excited state.
Diagrams (a)-(c) correspond to the usual
meson-exchange current
\ba
&&\langle ph^{-1}|{\hat j}_{\mu}^{MEC}|F\rangle
= -\frac{m f^2F_1^{(V)}
i\varepsilon_{3ab}}{V^2 m_\pi^2\sqrt{E_{\np} E_{\nh}}} 
\sum_{\nk}\frac{m}{E_{\nk}}
              \overline{u}(\np)\tau_a\tau_b
\left\{ \frac{(\Kbar-m)\gamma_\mu}{(P-K)^2-m_\pi^2}
\right.
\nonumber\\
&&
\left.
      + \frac{\gamma_\mu (\Kbar-m)}{(K-H)^2-m_\pi^2} 
- 2m
\frac{(Q+2H-2K)_\mu(\Kbar-m) }{[(P-K)^2-m_\pi^2]
         [(K-H)^2-m_\pi^2]}
\right\}u(\nh)
\ea
where the electromagnetic field is coupled with the pionic current ($V$ in the
volume enclosing the system and $f$ is the $\pi$-nucleon coupling constant);
diagrams (d)-(g) contain an intermediate virtual nucleon, 
described by the Feynman propagator $S_F$, and give rise 
to the so-called correlation current 
($\Gamma^\mu=F_1\gamma^\mu+iF_2\sigma^{\mu\nu}
Q_\nu/2m$ being the single-nucleon e.m. current):
\ba
&&
\langle ph^{-1}|{\hat j}_\mu^{C}|F\rangle
=           -\frac{1}{2m} \frac{m f^2}{V^2 m_\pi^2 \sqrt{E_{\np} E_{\nh}}}
             \sum_{\nk}\frac{m}{E_{\nk}}
              \overline{u}(\np) \left\{ \tau_a\gamma_5
              \frac{\Pbar-\Kbar}{(P-K)^2-m_\pi^2} 
              \right.
\nonumber\\
&& (\Kbar + m) \left[    \tau_a\gamma_5(\Pbar-\Kbar)
              S_F(H+Q)\Gamma_\mu(Q)
            + \Gamma_\mu(Q)S_F(K-Q)
              \tau_a\gamma_5
             \right. \nonumber\\
&&           \left. (\Pbar-\Kbar)
    \right]  + \left[    \tau_a\gamma_5(\Kbar-\Hslash)
              S_F(K+Q)\Gamma_\mu(Q)
            + \Gamma_\mu(Q)S_F(P-Q)
              \tau_a\gamma_5
\right.
\nonumber\\
&&\left.\left.(\Kbar-\Hslash)
    \right] 
              (\Kbar + m)\tau_a\gamma_5
              \frac{\Kbar-\Hslash}{(K-H)^2-m_\pi^2}
  \right\} u(\nh)\ .
\ea
The inclusion of the latter is crucial
for fulfilling gauge invariance: indeed it can be proved 
\cite{Ama01} that the relativistic MEC
and correlation currents satisfy current conservation,
{\it i.e.} 
\be
Q^\mu \langle ph^{-1}|{\hat j}_\mu^{MEC} + {\hat j}_\mu^C|F\rangle = 0\ ,
\ee
provided the same isovector electromagnetic form factor $F_1^{(V)}$ enters in
all of the currents.
\begin{figure}[h]
\begin{center}
\leavevmode
\epsfbox[130 500 530 680]{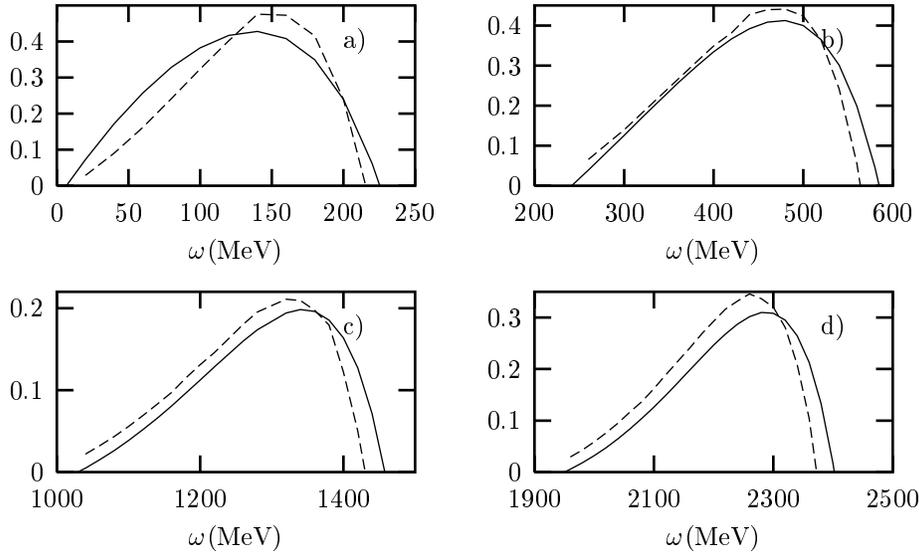}
\end{center}
\caption{
Longitudinal response versus $\omega$ including all first-order contributions
(dashed) compared with the free result (solid) at $q$=0.5 (a), 1 (b),
2 (c) and 3 (d) GeV/c.
Here and in all of the figures to follow the nucleus is $^{40}$Ca 
with $k_F$=237 MeV/c and the units are 
$10^{-1}$ MeV$^{-1}$ at $q$=0.5 GeV/c,
$10^{-2}$ MeV$^{-1}$ at $q$=1 GeV/c, 
$10^{-3}$ MeV$^{-1}$ at $q$=2 GeV/c and 
$10^{-4}$ MeV$^{-1}$ at $q$=3 GeV/c.
}
\end{figure}
\begin{figure}[h]
\begin{center}
\leavevmode
\epsfbox[130 500 530 680]{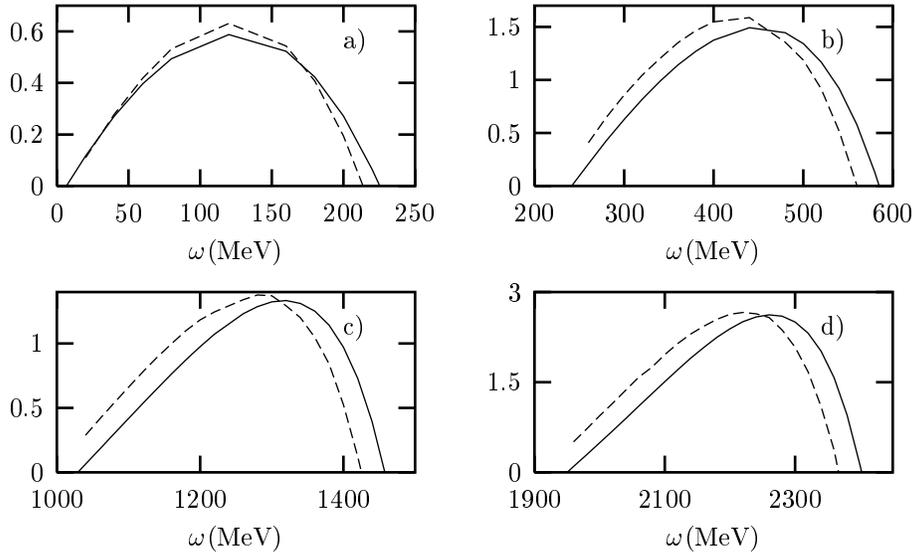}
\end{center}
\caption{
Same as Fig.~1 for the transverse response.
}
\end{figure}

The numerical results are displayed in Figs.~2 and 3, where 
the global responses in first order of perturbation theory are compared 
with the free ones for several momentum transfers.
The overall effect of the two-body currents appears 
sufficiently modest to justify our first-order treatment. 
Moreover, the softening at large $q$ appears to be common to both 
L and T channels, whereas at low $q$ the longitudinal response displays 
a hardening  that is absent in the transverse one.
Also evident is the almost vanishing of 
the two-body correlation contribution at the peak of the free responses.

A detailed analysis of the separate contributions of the diagrams
in Fig.~1
is presented in Refs.~\cite{Ama01,Ama01_1}. Here we summarize the main
results:
\begin{itemize}
\item
the MEC are almost irrelevant in the longitudinal channel, 
whereas teir contribution typically amounts to about 5--10\%
in the transverse one;
\item
the impact of the correlation current is substantial 
in both the  longitudinal and 
transverse responses, though it is actually dominant in the former, due
to the smallness of the isoscalar magnetic moment;
\item
the MEC and correlations effects
tend to cancel  in the transverse channel, especially for low values of
$q$, whereas for higher values of the momentum transfer the MEC dominate;
\item
the self-energy contribution arises from a delicate cancellation
of diagrams 1f and 1g, as already pointed out in Ref.~\cite{Bar93}
in a semi-relativistic context.
\end{itemize}
To stress the relevance of relativistic effects,
in Fig.~4 we compare the MEC contribution to the transverse
response with the non-relativistic calculation of~\cite{Ama94}
for $q=500$ MeV/c. It clearly  appears that, apart from the difference 
stemming from the relativistic kinematics which shrinks the response domain, 
the relativistic responses are about $30\%$ smaller than 
the non-relativistic ones, indicating that {\em relativity
plays an important role even for not so high $q$-values}.
\begin{figure}[h]
\begin{center}
\epsfig{file=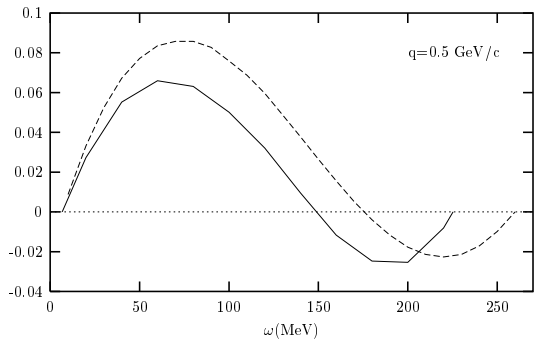,width=8cm}
\end{center}
\caption{
Transverse MEC contribution (solid) compared with the 
non-relativistic calculation of Ref.~{\protect\cite{Ama94}}
(dashed).
}
\end{figure}

We can now study the evolution of the MEC
and correlations with the momentum transfer and with the density.
Such dependences are clearly relevant in connection with analyses of scaling in
nuclei~\cite{yscaling}, both of first and second kinds.
From our results it emerges that:
\begin{itemize}
\item 
{\em scaling of first kind (independence of $q$) is attained at high 
momentum transfer};
\item
{\em scaling of second kind (independence of $k_F$) is broken through 
dependence roughly on the Fermi momentum squared for all values of $q$}.
\end{itemize} 
This point is best illustrated by displaying
the responses as functions of the scaling variable\cite{yscaling}
\be
 \psi \equiv\frac{E_0-m}{E_F-m}\ ,
\ee 
where $E_F$ is the Fermi energy and $E_0$ the minimum energy
needed for a nucleon inside the Fermi sphere to participate to the 
scattering process.  

In Fig.~5a the ratio $R^T_{MEC}/R^T_{free}$ is plotted for 
$q$ varying from 0.5 to 3 GeV/c: the relative MEC contribution 
decreases in going from 0.5 to 1 GeV/c, but then it rapidly saturates at
$q\simeq$1 GeV/c, where its value stabilizes, typically around 
10$\%$. Thus scaling of the first kind is satisfied.
In Fig.~5b the same ratio is plotted versus $\psi$ at $q$=1 GeV/c for 
$k_F$ varying from 200 to 300 MeV/c: it clearly appears that the relative 
MEC contributions grow with $k_F$ (attaining a value of about 20$\%$ at
$k_F=$300 MeV/c), indicating that the MEC content of $R^T$
increases with the denisty roughly as $k_F^2$, in contrast with the free 
response which decreases as $k_F^{-1}$. 
\begin{figure}[h]
\begin{center}
\leavevmode
\epsfbox[130  550 530 690]{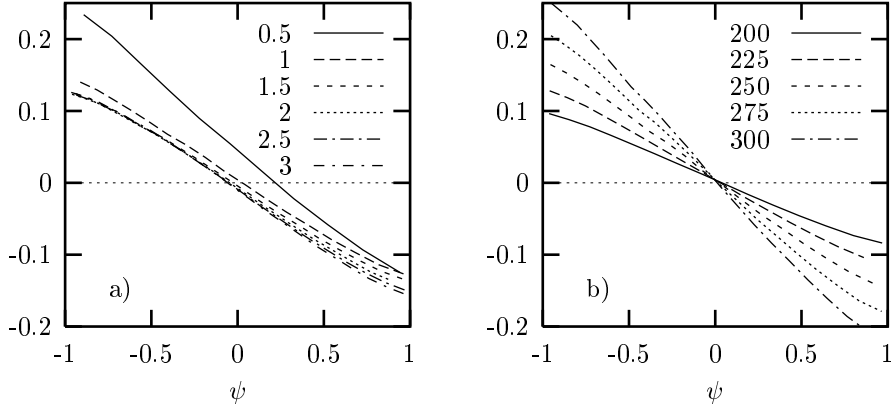}
\end{center}
\caption{
The ratio MEC/free in the transverse channel plotted versus 
$\psi$ at $k_F$=237 MeV/c for various values of $q$ (in GeV/c) in panel a and
for various values of $k_F$  (in MeV/c) at $q$=1 GeV/c in panel b.
}
\end{figure}
Similar results hold for the correlation contributions\cite{Ama01}.
We can thus conclude that the pionic effects in the reponse functions
{\em do not go away} as $q$ becomes very large and attain scaling of the first 
kind at momentum transfers somewhat above 1 GeV/c, while 
scaling of the second kind is badly violated.

\paragraph{Acknowledgments:}
I would like to thank
J.E. Amaro, J.A. Caballero, T.W. Donnelly and A. Molinari, who have 
collaborated in this work.


\begin{thebibliography}{99}
\bibitem{Ama98}
J.E. Amaro, M.B. Barbaro, J.A. Caballero, T.W. Donnelly, A. Molinari,
Nucl. Phys. {\bf A643} (1998) 349.
\bibitem{Ama99}
J.E. Amaro, M.B. Barbaro, J.A. Caballero, T.W. Donnelly, A. Molinari,
Nucl. Phys. {\bf A657} (1999) 161.
\bibitem{Alv01} L. Alvarez-Ruso, M.B. Barbaro,  T.W. Donnelly,  A. Molinari,
Phys. Lett. {\bf B497} (2001) 214.
\bibitem{Ama01}
J.E. Amaro, M.B. Barbaro, J.A. Caballero, T.W. Donnelly, A. Molinari,
nucl-th/0106035, to be published in Nucl. Phys. {\bf A}.
\bibitem{Ama01_1}
J.E. Amaro, M.B. Barbaro, J.A. Caballero, T.W. Donnelly, A. Molinari,
nucl-th/0107069, submitted to Nucl. Phys. {\bf A}.
\bibitem{Bar93}
W.M. Alberico, M.B. Barbaro, A. De Pace, T.W. Donnelly, A. Molinari,
Nucl. Phys. {\bf A563} (1993) 605.
\bibitem{Ama94} J.E. Amaro, G. Co', A.M. Lallena, 
Int. J. Mod. Phys. {\bf E3} (1994) 735.
\bibitem{yscaling}
T.W. Donnelly, Ingo Sick, Phys. Rev. Lett. {\bf 82} (1999) 3212 and
Phys. Rev. {\bf C60}:065502 (1999); 
W.M. Alberico, A. Molinari, T.W. Donnelly, E.L. Kronenberg,
J.W. Van Orden, Phys. Rev. {\bf C38} (1988) 1801.
\end{thebibliography}
\end{document}